%% file: main.tex
\documentclass[sigconf]{acmart}

\AtBeginDocument{%
  }

\copyrightyear{2026}
\acmYear{2026}
\setcopyright{cc}
\setcctype{by}
\acmConference[ICCAD '26]{IEEE/ACM International Conference on Computer-Aided Design}{November 08--12, 2026}{San Jose, CA, USA}
\acmBooktitle{IEEE/ACM International Conference on Computer-Aided Design (ICCAD '26), November 08--12, 2026, San Jose, CA, USA}
\acmDOI{10.1145/3831252.3834137}
\acmISBN{979-8-4007-2873-0/2026/11}

\usepackage{tabularx}
\usepackage{makecell}
\usepackage[table, dvipsnames]{xcolor}
\usepackage{textcomp}
\usepackage{enumitem} 
\usepackage{booktabs}   
\usepackage{siunitx}    
\definecolor{highlight}{HTML}{E0F2F1}

\newcommand{\AM}[1]{\textcolor{black}{#1}}

\begin{document}

\title{Adapting DNNs to Embrace Voltage Scaling Induced Errors via Accumulation-Aware Quantization 
}

\author{Alexis Maras}
\email{amaras@microlab.ntua.gr}
\orcid{0009-0001-4508-6384}
\affiliation{%
  \institution{National Technical University of Athens}
  \city{Athens}
  \country{Greece}
}

\author{Panagiotis Chaidos}
\email{pchaidos@microlab.ntua.gr}
\orcid{0009-0005-0165-5142}
\affiliation{%
  \institution{National Technical University of Athens}
  \city{Athens}
  \country{Greece}
}

\author{Georgios Alexandris}
\email{galexandris@microlab.ntua.gr}
\orcid{0009-0004-5900-9090}
\affiliation{%
  \institution{National Technical University of Athens}
  \city{Athens}
  \country{Greece}
}

\author{Dimitrios Soudris}
\email{dsoudris@microlab.ntua.gr}
\orcid{0000-0002-6930-6847}
\affiliation{%
  \institution{National Technical University of Athens}
  \city{Athens}
  \country{Greece}
}

\author{Sotirios Xydis}
\email{sxydis@microlab.ntua.gr}
\orcid{0000-0003-3151-2730}
\affiliation{%
  \institution{National Technical University of Athens}
  \city{Athens}
  \country{Greece}
}

\renewcommand{\shortauthors}{Alexis Maras et al.}

\input{abstract}


\maketitle

\input{introduction}
\input{related_work}
\input{ppa_analysis}
\input{error_modeling}
\input{methodology}
\input{evaluation}
\input{conclusions}

\bibliographystyle{ACM-Reference-Format-num}
\bibliography{references}

\end{document}

%% file: abstract.tex
\begin{abstract}

The pervasive deployment of Deep Neural Networks (DNNs) across the computing spectrum has brought growing attention to energy efficiency, alongside raw performance. While voltage scaling is a potent knob for power reduction, its application in next-generation nodes is hampered by severe throughput degradation and susceptibility to process variation. 
This work aims to 
\AM{address these energy-reliability trade-offs} by introducing a hardware-software co-design methodology, based on the observation that timing-robust bit ranges within the accumulator of a systolic array vary significantly across different voltage-frequency operating points.
At the software level, we propose an Accumulation-Aware Training approach that incorporates this information as a first-order optimization constraint during training. By integrating mixed-precision quantization with unstructured pruning, we effectively constrain the model’s weight distribution to avoid activating the hardware's sensitized high-rank carry chains. At the architectural level, we complement this with selective MSB-masking and fine-grained clock-gating to minimize unnecessary dynamic switching activity. Furthermore, we employ a layer-wise Dynamic Frequency Scaling strategy to further adapt to the accumulator's reduced dynamic range, based on the numerical sensitivity of each layer. \AM{We experimentally evaluate the proposed solution on a deep scaled 2nm technology node and demonstrate that, when operating at a low-voltage frontier, our co-optimized post-layout architecture delivers substantial energy and energy-delay 
product gains of 39\% and 23\% respectively, relative to the nominal voltage implementation, while maintaining accuracy degradation within 1\%.}

\end{abstract}

%% file: introduction.tex
\section{Introduction}

\par The widespread adoption of AI across the entire computing spectrum, from hyperscale data centers to resource-constrained edge devices, has shifted the design paradigm from raw performance \cite{tpu, dip} toward energy efficiency \cite{silvano_survey, ieeeelia}. 
Traditionally, DNNs have been adapted to support reduced-precision arithmetic formats without significantly compromising inference accuracy or output quality \cite{low_pr1, low_pr2}. Quantization offers substantial architectural advantages by effectively reducing the memory footprint and alleviating traffic on the global interconnect, while leveraging the increased computational density of smaller arithmetic units \cite{mpq_intro, mpq_intro2}. However, the efficacy of quantization can sometimes be bounded, particularly when adopting aggressive sub-byte precisions that can lead to severe accuracy degradation \cite{matryoshka}. Consequently, several works have explored hardware-software co-design solutions, developing flexible hardware components tailored for low-bitwidth operations alongside software frameworks to maintain model robustness \cite{mixgemm}. 

\par Meanwhile, voltage scaling has emerged as one of the most effective design techniques for improving the energy-efficiency of DNN accelerators; specifically, near-threshold voltage (NTV) represents an optimal operative regime from an energy standpoint \cite{ntv_optimal}. While a reduction in supply voltage significantly reduces power consumption, it also introduces the risk of erroneous outputs due to critical paths failing to meet timing requirements \cite{razor}. Although DNN algorithms exhibit an inherent tolerance to traditional approximation techniques \cite{axrblocks}, the propagation of timing errors results in a catastrophic accuracy degradation, particularly in architectures featuring massive parallelism, such as large-scale MAC arrays \cite{thundervolt_opt}. Consequently, strict operating-frequency management is a prerequisite for reliable inference \cite{freq_adj}. This constraint implies that aggressive voltage scaling is inherently coupled with diminished throughput. Finally, under aggressive voltage scaling operation, circuits showcase increased sensitivity to manufacturing variations that can result in high deviations in drive current and propagation delay, further exacerbating the likelihood of timing violations and complicating the pursuit of a robust, low-power design \cite{varius_ntv, ntv_synthesis}.

\par Previous research exploring voltage scaling to optimize the energy efficiency of DNN accelerators has largely failed to capitalize on the synergistic benefits offered by reduced-precision arithmetic formats and their corresponding hardware units \cite{xtpu, xnvdla}. Instead, these works typically focus on architectural modifications designed to predict \cite{greentpu} or correct \cite{thundervolt_opt} erroneous results occurring within logic paths and memory blocks \cite{erdcim, sram_boost}. By treating voltage scaling and quantization as independent optimization goals, these approaches often overlook the opportunity to co-optimize the hardware for both bit-width reduction and supply voltage minimization. Furthermore, existing methodologies \cite{gavina, erdcim, sram_boost, xtpu, xnvdla, mac_vt} frequently fail to account for or co-optimize their systems against the effects of process variability. By overlooking the interplay between voltage-induced delays and manufacturing-induced fluctuations, these designs leave a significant reliability gap, particularly in NTV conditions where the lack of variation-aware optimization can lead to unpredictable system failures \cite{vt_var}. \AM{\textit{Given these limitations, a critical question arises: can we balance these reliability-energy trade-offs by proactively manipulating the model's parameters?}}

\par Addressing this question is the primary motivation of this work. To achieve this, we first conduct a thorough analysis of a systolic array architecture at a 2nm technology node to quantify the benefits and limitations of low-voltage operation. Based on the observations of this characterization, we leverage the bit-order dependency of timing failures, which are structurally localized within the high-order carry chains of the accumulators, to implement a Variation-Resilient Architecture featuring selective MSB-masking and fine-grained clock-gating, eliminating dynamic switching power in the corrupted bit range. Furthermore, our methodology incorporates Accumulation-Aware Training (AAT) \cite{a2q}, utilizing Mixed-Precision Quantization and unstructured pruning to mathematically constrain the model’s weights, forcing intermediate products to reside strictly within the hardware's reliable "safe zone" (i.e., the bit-range within the accumulators, where timing violations are avoided). Finally, acknowledging that overly restrictive accumulator constraints can trigger accuracy loss \cite{wrapnet}, we introduce a layer-wise Dynamic Frequency Scaling (DFS) strategy to adaptively extend the accumulators' "safe zone" for sensitive layers without necessitating a global frequency reduction.

\textbf{Our main contributions in this work are the following:}

\begin{enumerate}[topsep=0pt]
    \item We provide a comprehensive evaluation of the performance and power trade-offs for systolic architectures at the 2nm GAAFET node, characterizing the process variation effects that dictate the frequency-scaling limits at both the nominal and low-voltage frontiers.
    \item We propose a software-hardware co-design methodology that exploits the \AM{localization of timing violations within the accumulators of the design}, even under high process variation conditions. This architecture employs selective MSB-masking, fine-grained clock-gating, \AM{and DFS to create an adaptive} hardware "safe zone".
    \item We introduce a \AM{flexible} AAT methodology, enforced with Mixed-Precision Quantization and unstructured pruning to guarantee the avoidance of accumulation overflow beyond the reliable limits of operation.
\end{enumerate}

\par By treating the hardware’s timing slack as a first-order constraint during optimization, we demonstrate that both conventional CNN architectures and Transformer-based models can maintain functional integrity at a low voltage regime of 0.5V, while on average, improving energy consumption and EDP by 39\% and 23\% respectively, \AM{compared to the nominal (0.7V) baseline implementation}.

%% file: related_work.tex
\section{Related Works \& Limitations}
\label{sec:related}

\par A significant body of research has investigated aggressive voltage scaling as a primary strategy for enhancing system energy efficiency. A prominent research direction focuses on voltage overscaling, in which supply voltage is reduced within computational and logic data paths without a proportional decrease in operating frequency, which would otherwise compromise throughput \cite{xtpu, gavina, greentpu}. By maintaining high clock frequencies during near-threshold operation, these designs achieve substantial energy gains while intentionally operating within the subcritical timing region. To mitigate the resulting timing violations, systems typically adopt one of the following paradigms to manage transient hardware faults: Timing Error Detection \& Timing Error Propagation \cite{gavina}. In the first approach, systems employ detection mechanisms during runtime to identify delayed signals. Once a violation is detected, the erroneous outputs are suppressed, often by masking the results to zero \cite{thundervolt_opt}, relying on the network's stochastic nature to maintain functional correctness, or the operation is re-computed by dynamically scaling the voltage to a safer operating region \cite{greentpu}. Alternatively, in the second approach, selective voltage scaling is applied to specific hardware structures, targeting either the LSBs of multipliers and adders \cite{xnvdla, gavina}, or dynamically scaling the supply voltage during inference on components that compute individual neurons \cite{xtpu, mac_vt}. This second approach ensures that the resulting computational inaccuracies are confined to elements whose contribution to the overall inference accuracy is statistically negligible.

\begin{figure}[t] 
    \centering \includegraphics[width=\columnwidth] {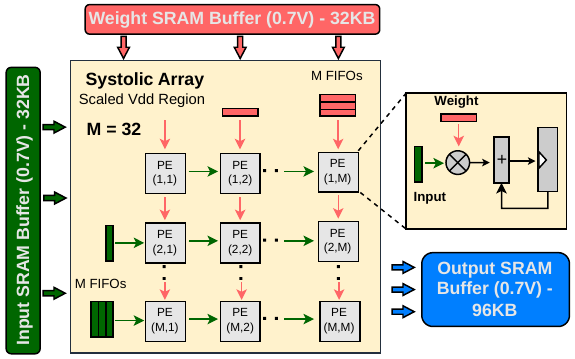} 
    \caption{Overview of the examined output stationary systolic array, along with the considered scaled Vdd region.} 
    \label{fig:sys_array} 
    \Description{Overview of Baseline Architecture}
\end{figure}

\par Existing studies on aggressive voltage scaling have largely overlooked the nuanced reliability implications arising from the interaction between numerical precision and underlying device physics. Most prior research \cite{xtpu, xnvdla, thundervolt_opt} treats hardware as a monolithic entity, failing to capitalize on how varying the bit-precision of weights and activations fundamentally alters the timing-slack and error profiles of the MAC logic. Moreover, as architectures scale toward advanced nodes, localized process variations, particularly fluctuations in threshold voltage \cite{freq_adj}, create a non-uniform reliability landscape across critical design components. Current literature does not explore how these physical variations sensitize specific logic paths, such as high-order carry chains, or whether this sensitivity can be mitigated by adjusting the numerical dynamic range. By failing to integrate device-level timing guardbanding with variation-aware training, existing approaches miss critical opportunities to harden accelerators against the bit-specific glitches inherent in low-voltage regimes, leaving a significant gap between energy-efficient scaling and the accuracy requirements of modern AI workloads \cite{vt_var}.

\par To address the aforementioned shortcomings, we propose a cross-layer co-design methodology that moves beyond generic error modeling by anchoring DNN optimization in the physical timing slack. By rigorously studying the underlying systolic array architecture under low-voltage scenarios and the effects of process variation, we identify a localized failure mechanism within the accumulator’s high-order bits. To mitigate their effect, we propose Accumulation-Aware Training, coupled with mixed-precision quantization and unstructured pruning to mathematically steer the model's activations into a deterministic hardware-compliant window. Finally, we employ a layer-wise DFS scheme to adaptively extend the accumulators' "safe zone", specifically for layers that showcase a higher sensitivity to reduced widths.

%% file: ppa_analysis.tex
\section{Low-Voltage TPU \& the Impact of Process Variations}
\label{sec:analysis}

\par Since the primary objective of this work is to optimize the energy consumption through voltage scaling while mitigating the associated reliability risks, we first analyze the energy and delay profile of the targeted micro-architecture: an output-stationary systolic array, due to its widespread adoption and proven efficacy in both academic research and industry-standard AI accelerators, such as the Google TPU \cite{tpu}. This architecture is renowned for its high degree of modularity and scalability, and it additionally provides a regular hardware substrate for characterizing the effects of aggressive voltage scaling, establishing a rigorous Power Performance Area (PPA) baseline to quantify our optimization gains. While prior works have explored the reliability-energy trade-offs of voltage scaling within the SRAM buffers \cite{minerva, erdcim, sram_boost}, these strategies have been evaluated on more mature nodes (22nm and above) and have yet to be adapted to the exacerbated variability of sub-5nm regimes \cite{sram_variability}. This study pivots from memory-centric mitigation to examine the impact of low-voltage operation on the computational datapaths of a systolic array, showcasing its potential and effectiveness.

\subsection{Experimental Setup}

\par Our analysis centers on a micro-architecture employing a $32\times32$ systolic MAC array with 8-bit precision for weights and activations and 24-bit accumulators; a configuration tailored for edge applications, targeting a clock of 2GHz as our baseline \AM{(an overview is shown in Figure \ref{fig:sys_array})}. The architecture is implemented using the N2 PDK provided by IMEC \cite{imec2025nanoic}. To establish an industry-standard performance baseline, arithmetic units are implemented via the Synopsys DesignWare library. For the physical implementation, we employed the Cadence back-end suite, utilizing Innovus for place-and-route (PnR) and Tempus for timing sign-off. 

\par To establish a quantitative foundation for our work, we characterize the power and delay profile of the 2nm systolic array across four distinct voltage regimes. These include a high-performance \qty{0.9}{\V} region \AM{(to match the latency of reading from the SRAM buffers to the FIFOs of the array)}, the nominal \qty{0.7}{\V} sign-off point, and two aggressive low-voltage frontiers at \qty{0.5}{\V} and \qty{0.4}{\V}. All regimes are evaluated at the Slow-Slow (SS) process corner to capture the worst-case leakage and timing scenarios inherent in the N2 node. Our timing analysis accounts for manufacturing variability by modeling the statistical extremes of gate-length, equivalent dielectric thickness, fin height, and work-function fluctuations. In contrast to more complex systems, such as general-purpose CPU cores, where the deepest critical paths may be activated only by infrequent instruction sequences \cite{cpu_ntv, dfb}, systolic arrays exhibit a high degree of structural regularity. As a result, the "worst-case" timing scenario is not a rare occurrence but the operational norm. This regularity implies that any process-induced shift that compromises a critical carry-chain in a MAC unit will manifest as a persistent error rather than a transient anomaly. Finally, given that the impact of variability on the threshold voltage ($V_{th}$) is particularly dominant in 2nm GAAFETs, we specifically present the resulting deviations in delay when accounting for these stochastic fluctuations \AM{(Figures \ref{fig:corner_analysis} and \ref{fig:error_heatmap})}. At the lower voltage regimes, even minor $V_{th}$ shifts translate into exponential deviations in propagation delay \cite{vt_var, varius_ntv}. By quantifying these deviations, we can map the erosion of setup slack.

\subsection{Timing Analysis} 
\label{subsec:timing_analysis}
\begin{figure}[t] 
    \centering \includegraphics[width=\columnwidth] {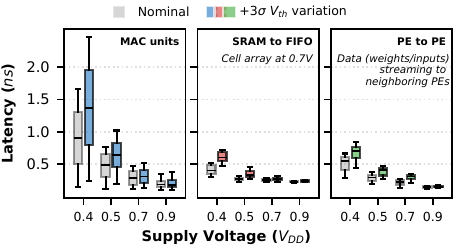}
    \caption{Delay Distribution in major components of the Systolic array design at different $V_{dd}$ corners and under the effects of Threshold Voltage Variation.}
    \label{fig:delays_boxplots} 
    \Description{Delay/Power Corner Analysis}
\end{figure}

\par \AM{We quantify the critical paths of the post-PnR designs by examining the timing reports in each operating condition. As shown in Figure \ref{fig:delays_boxplots}, our analysis verifies that the critical paths of the design reside within the combinational logic of the accumulator's adder stages. This holds even under a heterogeneous-voltage scheme, in which the SRAM memory buffers are maintained within a stable 0.7V range to ensure read/write stability (250 psec access latency). While the memory often dictates the Vmin of the system, the logic-path delay within the PE array becomes the primary bottleneck for high-frequency operation as voltage scales, due to the diminished drive strength of the GAAFET devices at NTV conditions. The deep carry-chain paths and sign-extension logic within the MAC units not only exhibit the highest sensitivity to voltage-induced slack erosion, but more importantly, their long logic paths are more sensitive to threshold voltage fluctuations, causing cumulative timing delays that disproportionately affect the high-order bits. In contrast, the shorter register-to-register paths, such as the interconnects between PEs and the SRAM-to-FIFO interfaces that are dominated by repeaters and low-depth buffering logic with significantly higher timing slack, remain relatively robust, therefore leaving the MAC unit's accumulation chain as the primary bottleneck.}

\par
\begin{figure}[t] 
    \centering \includegraphics[width=\columnwidth] {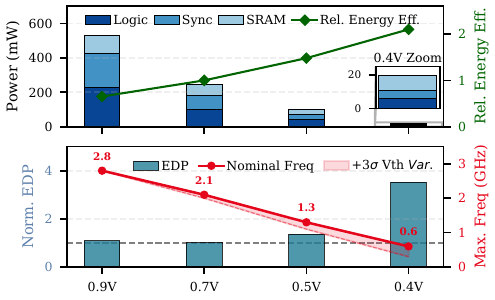} 
    \caption{Comparative analysis of the Systolic Array across different supply \& threshold voltage conditions.} 
    \label{fig:corner_analysis} 
    \Description{Delays per component in the design}
\end{figure}

 This performance degradation is further illustrated in Figure \ref{fig:corner_analysis}, as scaling from high-performance 0.9V and nominal 0.7V regimes to the near-threshold frontier fundamentally shifts the systolic array's temporal constraints. While the array achieves 2.8 GHz at 0.9V and 2.1 GHz at 0.7V, the transition to lower voltages triggers a non-linear performance collapse. At 0.5V, the nominal frequency drops to 1.3 GHz, and at 0.4V, it plummets to just 0.6 GHz. The introduction of $V_{th}$ variations further constricts the frequency envelope (the shaded red region); to mitigate timing violation risks, the maximum operational frequency must be throttled down to 1 GHz and 400 MHz for the 0.5V and 0.4V corners, respectively. \AM{Since the critical paths reside within the MAC units, localized variations that could potentially affect even one unit randomly across the die could potentially bottleneck the performance of the synchronous array.}

\subsection{Power \& EDP Analysis}

\par We complement our characterization by explicitly capturing the complex physical interactions of the full 2nm systolic array, accounting for the capacitive overhead of the global interconnect and clock distribution network. As illustrated in the power breakdown (Figure \ref{fig:corner_analysis}), scaling the supply voltage from the nominal 0.7V sign-off point to the 0.5V near-threshold regime yields a significant reduction in total power, dropping from 254 mW to approximately 90 mW, resulting in a 61\% drop. At the extreme 0.4V frontier, the power floor further collapses to roughly 20 mW. Conversely, pushing the array to 0.9V, to match the peak access latency of the SRAM buffers, leads to a significant power surge to 532 mW, more than doubling the dissipation of the nominal case. These substantial power reductions are not solely a function of VDD scaling; they are also heavily attributed to the necessary reduction of the clock frequency to avoid timing violations under the $V_{th}$ variations analyzed previously. Across the entire voltage range, the synchronization (Sync) overheads (registers) and the clock tree network remain a persistent energy bottleneck, contributing roughly 35\% of the total power. We also observe that the relative power contribution of SRAM accesses becomes more prevalent at lower supply voltages, as the memory cell arrays must be maintained at a higher VDD to ensure stability while the surrounding logic scales. Since clocking and logic-switching and the wire-load capacitances dominate the energy footprint in this technology node \cite{provet}, voltage scaling remains a technique capable of simultaneously attenuating both.

\par Regarding energy efficiency, the benefits of near-threshold operation are evident: the TOPs/W of the accelerator increase by factors of 1.5$\times$ and 2.0$\times$ as the supply voltage is reduced to 0.5V and 0.4V, respectively. However, this primary motivation must be rigorously balanced against \AM{the associated performance losses, by focusing on how the EDP metric scales in each operating point}. At the high-performance 0.9V point, despite achieving a peak frequency of 2.8 GHz, the massive power overhead leads to a sub-optimal 1.09$\times$ higher EDP relative to the 0.7V baseline. This suggests that "over-volting" for raw throughput can yield diminishing returns. This efficiency gap becomes even more pronounced in the low-voltage regimes due to the frequency collapse. At 0.5V, the required frequency reduction to 1.0 GHz (under worst-case variation) results in an EDP 1.45$\times$ higher than the 0.7V baseline, while at 0.4V, EDP skyrockets to 3.5$\times$ the nominal, indicating that the performance penalty associated with NTV can outweigh the energy savings. 

%% file: error_modeling.tex
\section{Error Propagation Analysis}

\begin{figure}[t] 
    \centering \includegraphics[width=\columnwidth] {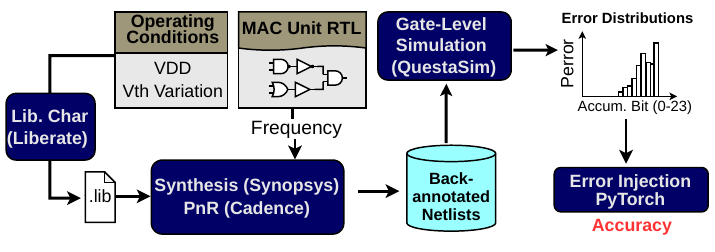} 
    \caption{End-to-end DNN evaluation methodology for examined operating conditions and extracted error distributions.} 
    \label{fig:dnn_eval} 
    \Description{Gate-Level Error Evaluation Methodology}
\end{figure}

\textbf{Bit-Level Error Distributions}: \AM{We elaborate on the fault modeling of the timing violations by performing a detailed timing characterization, by following the procedure shown in Figure \ref{fig:dnn_eval}. In our analysis, the $V_{th}$ variability (considering a 3$\sigma$ deviation from the nominal value as in previous works \cite{ntv_synthesis})} and target supply voltages are provided as inputs to the Library Characterization stage, executed with Cadence Liberate. This ensures that the resulting \textit{.lib} files inherently encapsulate the process-variation effects that will subsequently drive the physical implementation of the MAC units \AM{that these were identified as the primary bottlenecks of our design in low-voltage regimes (Section \ref{subsec:timing_analysis})}. By performing Gate-Level Simulations with Siemens' QuestaSim on the back-annotated netlists, we map the distribution of the \AM{erroneous} timing violations across the \AM{accumulators' bits}.

\par These simulations reveal a distinct correlation between the $V_{th}$ fluctuations and bit-specific error probabilities, particularly in the high-order carry chains. More specifically, as illustrated in Fig. \ref{fig:error_heatmap}, when variability in the $V_{th}$ value is not taken into account, the \AM{error distribution (flipped bit value)} becomes much more localized. In these scenarios, timing violations are almost exclusively confined to the 7 MSBs of the 24-bit accumulator (further verifying the observations from \cite{thundervolt_opt}). In contrast, under conditions of higher $V_{th}$ variability (i.e., a worst-case but realistic scenario), the probability of timing errors is elevated even across lower-order bits, with violations beginning to manifest as early as the 12th bit position. \AM{This broader error profile occurs because threshold fluctuations slow down the initial carry-generation stages, causing a "ripple effect" of delayed transitions that propagate through the entire accumulation path and fail to settle within the clock cycle.}

\noindent\textbf{Does bit-width matter?} Intuitively, one might assume that INT4 operations, by virtue of their reduced word length, would exhibit a lower probability of timing failure compared to INT8 operations. However, our back-annotated timing analysis reveals that this is not the case, as INT4 and INT8 parameters demonstrate a nearly identical probability of output corruption. This phenomenon stems from the physical nature of timing failures, as they typically manifest as transient glitches captured at the rising edge of the clock. Multiple logic paths within the MAC unit are activated simultaneously but fail to settle before the latching edge. Consequently, all bits, including the higher-order sign-extension ones, may momentarily take a value of "1" or "0" as the logic waves propagate through the carry chains. Because these glitches are a byproduct of the total switching activity within the 24-bit accumulator, the reduced bit-width of an INT4 operand does not provide a safety margin, and the internal carries and accumulation logic still trigger the same unstable switching behavior in the MSBs. Nevertheless, low bit-width parameters can be utilized as a means of algorithmic regulation rather than physical timing mitigation. By aggressively reducing the precision of weights and activations, we can effectively constrain the numerical magnitude of the accumulated results. This ensures that the final values remain within a "safe zone" of the accumulator, allowing us to extract the correct result by masking the corrupted high-order bits that fall outside this intended range.

\noindent\textbf{End-to-End DNN Evaluation}: Low-voltage operation introduces a distinct severe error modality: high-magnitude, data-dependent transients localized in the accumulation stage. To evaluate this, we injected such errors, sampled from our gate-level characterization, into the accumulated results of the intermediate feature maps of a ResNet-50 model. Even if timing violations manifest in only a single MAC unit (the specific location of which varies from die to die), the probability of failure, shown in Figure \ref{fig:error_heatmap}, is high enough to corrupt the generated feature maps, causing Top-1 accuracy to collapse to 0\%. Such catastrophic failures underscore that allowing unmitigated timing-error propagation in systolic architectures is prohibitive without hardware-software countermeasures \cite{thundervolt_opt}.

%% file: methodology.tex
\section{Reliability-Aware TPU Architecture \& Variation Adaptive Fine-Tuning}

\par To recover the functional accuracy and energy-delay efficiency of the systolic array at the 0.5V frontier \AM{(we select this voltage level due to its potential to match the performance of the 0.7V operating point)}, we propose a cross-layer framework that harmonizes hardware-level resilience with software-level adaptation. \AM{Our methodology is predicated on the insight that near-threshold timing violations are not stochastic noise; rather, they are structurally localized within specific bit-positions.} By treating the hardware’s physical limitations as a first-class constraint during the DNN fine-tuning phase, we move beyond generic error tolerance toward a "design-for-variation" paradigm. This approach is structured into three synergistic pillars: (i) an architectural modification that employs selective clock-gating and MSB-masking to create a deterministic hardware "safe zone," (ii) \AM{Accumulation-Aware Training \cite{a2q} to fine-tune the DNNs to operate within truncated accumulator bit-widths}, (iii) Mixed-Precision Quantization (MPQ) along with unstructured pruning to statistically shift the accumulation magnitude away from sensitized logic paths. 
\begin{figure}[t] 
    \centering \includegraphics[width=\columnwidth] {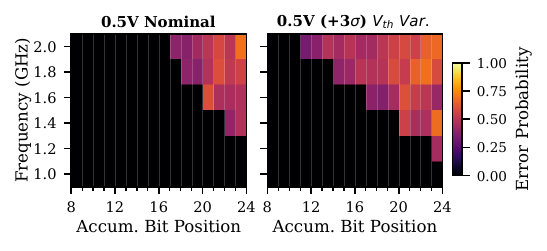} 
    \caption{\AM{Error Distribution of the accumulator within MAC units at 0.5V as we scale the frequency, along with the impact of threshold voltage variability on the error probability.}} 
    \label{fig:error_heatmap} 
    \Description{Error Distribution of the MAC accumulator.}
\end{figure}

\subsection{MSB-Masking and Selective Clock-Gating}

\par As established in our timing characterization, process variations at the low-voltage primarily sensitize the sign-extension logic and high-order carry chains, while the lower bits of the 24-bit accumulator remain robust. We can capitalize on this spatial locality by introducing a "Safe-Mode" hardware abstraction at the PE level. This architectural modification consists of programmable masking logic situated at the accumulator's output, acting as a hardware-level 'firewall' that truncates the sensitized MSBs, since the content of these high-order bits is considered numerically redundant in our optimized dataflow. It is important to note that maintaining the full 24-bit width is essential for architectural flexibility; removing these bits entirely would impose excessively rigid constraints on the model's dynamic range, which is necessary for high-precision tasks/layers. To further capitalize on this localized sensitivity, we implement fine-grained selective clock-gating for the accumulator's registers and the logic associated with the paths that lead to the faulty MSBs. Given that the clock distribution network and synchronization overhead account for roughly 35\% of the total array power, gating the clock signal to the sensitized flip-flops provides an extra opportunity to improve energy efficiency. Physically, this effectively "freezes" the unreliable logic paths, eliminating the dynamic switching power of the high-order bits that the software is trained to ignore. The implementation of these minimal mechanisms incurs a minimal area penalty of 1.7\% in the total cell area.

\par \AM{Finally, since extremely narrow accumulator bit-widths can lead to significant accuracy loss \cite{wrapnet, a2q}, particularly in layers with high numerical variance that cannot be fully mitigated even after Accumulation Aware Fine-Tuning (discussed in Section \ref{subsec: finetuning}), we propose a Layer-Wise DFS strategy. For these layers, the system will operate at a lower frequency} thereby effectively expanding the accumulator's "safe zone", as shown in Figure \ref{fig:error_heatmap}. \AM{This ensures that accuracy is prioritized for high-saliency layers without necessitating a clock decrease for the entire network, thereby preserving the overall energy efficiency We should note that the overheads associated with this DFS scheme are negligible, since the frequency settings are gathered offline during the fine-tuning phase, and the hardware performs a low-latency transition between pre-characterized clock domains at (coarse-grained) layer boundaries.}

\par The quantitative impact of this DFS strategy is detailed in Table \ref{tab:rel_edp_analysis}, which highlights four critical operating regimes at 0.5V. At 1.0 GHz, the architecture operates in the full-precision zone, providing a 24-bit safe accumulator width for layers requiring maximum numerical range, albeit with a 45\% EDP overhead relative to the 0.7V baseline. Scaling to 1.4 GHz represents the iso-EDP point, where the 19-bit safe-mode matches the efficiency of nominal-voltage operation while maintaining 2nm scalability. For aggressive optimization, 1.8 GHz serves as a strategic midpoint, providing a 14-bit safe zone with a significantly reduced normalized EDP of 0.63. Finally, at 2.0 GHz, the system achieves iso-performance relative to the nominal baseline; in this regime, although the safe zone is limited to 11 bits, the hardware delivers a 51\% reduction in EDP. By dynamically transitioning between these modes, we ensure that each layer utilizes the highest possible frequency that its learned weight distribution can tolerate without inducing timing-driven corruption.

\input{Tables/edp_comparison}

\subsection{Accumulation-Aware Quantization}
\label{subsec: finetuning}

\par To ensure the model's accuracy remains under the hardware's MSB-masking, we must guarantee that the accumulated results do not overflow the deterministic bit-width of the MAC units. Let $P^*$ denote the minimum accumulator bit-width at a given time step required to guarantee overflow avoidance for a given dot product. The most conservative method to calculate $P^*$ considers the width of the dot product operands, given that inputs $\tilde{x}$ and weights $\tilde{w}$ are both signed and quantized to $N$ and $M$ bits respectively:

\begin{equation}
    P^* = \lceil \log_2(2^{\log_2(K) + N + M - 1}) + 1) \rceil
\end{equation}

\noindent This equation highlights that $P^*$ increases linearly with the bit-widths of the operands and logarithmically with the depth of the dot product. Another approach is to treat the accumulator bit-width as an independent dimension of the quantization design space. Specifically, when accumulating $\tilde{x}^T w$ into a signed $P$-bit accumulator, and assuming a zero-mean weight distribution ($\sum_i w_i = 0$), we can theoretically guarantee overflow avoidance by constraining the $\ell_1$-norm of the weights, by enforcing the following constraint\protect\footnote{\AM{The formal derivation is provided in \cite{a2q}}}:

\begin{equation}
    \|w\|_1 \le \frac{2^{P}-2}{2^{N}-1}
    \label{eq:l1norm}
\end{equation}

\par In contrast to the original work \cite{a2q}, we aim to meet these constraints by employing a specialized cross-layer strategy. Our approach combines a mixed-precision scheme with selective unstructured pruning to naturally lower the accumulation magnitude. Furthermore, we fine-tune the model considering an accumulator with an adaptive effective size. This is critical for our framework, as the accumulator's safe bit-width is not a static architectural constant, but is instead dynamically defined by the target operating frequency and the resulting timing slack at the 0.5V corner. By integrating this adaptive constraint into the training loop, we allow the model to converge even when accumulator constraints are exceptionally tight for some of the layers.

\AM{\subsubsection{\textbf{Algorithmic Range Control with Mixed-Precision and Pruning:}} To align the model's numerical behavior, we propose a dual-step optimization strategy that exploits the ability of DNNs to function under reduced bit-widths and high sparsity to alleviate the strict constraints of equation \ref{eq:l1norm}. By employing MPQ, specifically transitioning to sub-byte representations, we aim to truncate the dynamic range of the accumulated partial products while preserving the model's predictive capabilities \cite{marvin}. To complement this, we utilize unstructured pruning as a fine-grained mechanism for magnitude control. Unlike structural pruning, the higher sparsity potential of unstructured methods allows us to aggressively "thin out" weight tensors in numerically volatile layers.}

\subsubsection{\textbf{Integration within Training}}

\par To enforce the hardware-driven constraints during optimization, we incorporate the $\ell_1$-norm (equation \ref{eq:l1norm}) penalty directly into the loss function as a differentiable regularizer. This penalty is applied at a granular level to every hidden layer and aggregated into a global regularization term:

\begin{equation}
\mathcal{L}{reg} = \sum_{l} \sum_{i} R_{l,i}
\end{equation}

\noindent where $R_{l,i}$ denotes the regularization penalty for the i-th output channel in the l-th layer of the network. To maintain training efficiency and model stability, we adopt a selective fine-tuning strategy: we freeze the weights of all layers that satisfy the accumulator constraints following the MPQ assignment and initial pruning, and only fine-tune the weights of the "at-risk" layers that exceed the safe-mode threshold. The specific penalty R for a given weight vector w is defined based on the distance between its current magnitude and the theoretical overflow threshold:

\begin{equation}
    R = \max(\|w\|_1- \Phi, 0)
\end{equation}

\noindent where $\Phi$ represents the theoretical value required to guarantee overflow avoidance for the target bit-width (e.g., P=14). This formulation ensures that no penalty is incurred as long as the weights remain within the safe operational regime of the 2nm PE logic. Finally, we scale this regularization term by a constant scalar $\lambda$ \AM{to control the "stiffness" of the hardware constraint; a higher value forces the weights to converge strictly within the accumulator's "safe zone," while a lower value prioritizes the preservation of the model's predictive features. Similar to \cite{a2q} we set the value to $10^{-3}$ to balance the two objectives}:

\begin{equation}
\mathcal{L}{total} = \mathcal{L}{task} + \lambda \mathcal{L}_{reg}
\end{equation}

\noindent where $\mathcal{L}{task}$ is the standard task-specific loss (e.g., Cross-Entropy). By using this approach, gradient descent naturally pulls the weight distributions into a range compatible with the timing constraints.

\subsubsection{\textbf{Exploration Loop}}

\begin{figure}[t] 
    \centering \includegraphics[width=\columnwidth] {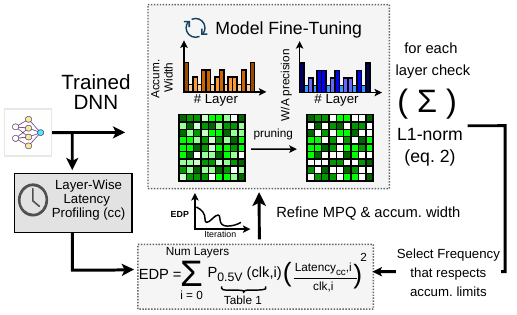} 
    \caption{\AM{Accumulation-Aware Exploration Flow. Iterative EDP optimization through layer-wise selection of weight and accumulator bit-widths, guided by configurations in Table \ref{tab:rel_edp_analysis}.} 
    } 
    \label{fig:methodology} 
    \Description{Fine-Tuning Methodology}
\end{figure}

\par Figure \ref{fig:methodology} summarizes our optimization and exploration loop, which treats the EDP as the primary optimization target. The loop begins by extracting the latency of each layer in the network for the architectures that we have defined in Section \ref{sec:analysis} with the use of ZigZag \cite{zigzag}. An initial MPQ (8-bit for both activations and weights for the first and last layer, while for the rest we start with 4-bits) configuration is selected, and the network is trained to satisfy our 1\% accuracy degradation threshold. If the model fails to reach this criterion, the framework identifies layers with the lowest latency contribution and selectively restores them to higher bit-widths, recovering predictive signal with minimal impact on the total system-level EDP. Once the accuracy target is met, we apply unstructured pruning to the weights to further reduce their numerical magnitude. Following pruning, we evaluate each layer against the $\ell_1$-norm boundary defined in Equation \ref{eq:l1norm}. For layers that satisfy this condition, the weights are frozen to guarantee hardware compatibility. Conversely, for layers that fail to meet the boundary, we apply our Accumulation-Aware Quantization methodology to force numerical compliance. If a layer remains a "bottleneck" even after AAT, the loop triggers a secondary "relief valve" through layer-wise DFS. \AM{Finally, the loop then proceeds by identifying the layers that currently have the highest relative EDP contribution and iteratively pushing them toward lower bit-widths and narrower accumulation widths.}

\begin{figure*}[t] 
    \centering 
    \includegraphics {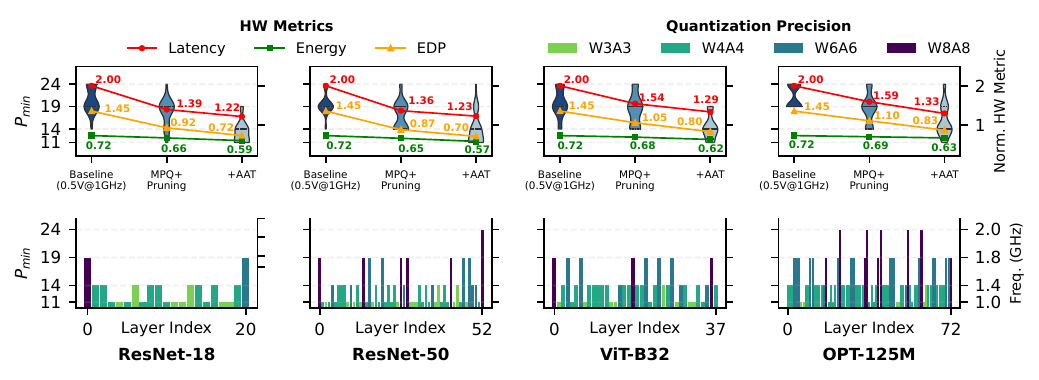} 
    \caption{Evaluation of hardware-aware optimization stages under 1\% loss accuracy constraint. The top row depicts the distribution of the minimum required accumulator bit-width ($P_{min}$), overlaid with normalized hardware metrics (Latency, Energy, and EDP) relative to the (INT8, 0.7V @ 2GHz) baseline. \AM{The bottom row illustrates layer-wise breakdowns of the final $P_{min}$ assignments, while the bar colors denote the quantization precision assigned to each layer, highlighting their correlation.}} 
    \label{fig:optim_results} 
    \Description{Methodology Results}
\end{figure*}

%% file: Tables/edp_comparison.tex
\begin{table}[t]
    \centering
    \setlength{\aboverulesep}{0pt} 
    \setlength{\belowrulesep}{0pt}
    \caption{Accumulator Reliability and Efficiency Metrics ($V_{DD} = \qty{0.5}{\V}$) after MSB masking and clock-gating}
    \label{tab:rel_edp_analysis}
    \renewcommand{\arraystretch}{1.05} 
    
    \begin{tabular}{@{} c c c c c @{}}
    \toprule
    \textbf{Freq.} & \textbf{Safe} & \textbf{Power} & \textbf{Norm. EDP} & \textbf{Efficiency} \\
    (GHz) & \textbf{Bits} & (mW) & (vs. \qty{0.7}{\V}) & (TOPS/W) \\ \midrule
    
    \rowcolor{green!25} 1.0 & 24 & 90  & 1.45 & 22.7 \\
    1.2 & 22 & 109 & 1.24 & 22.5 \\
    \rowcolor{green!25} 1.4 & 19 & 117 & 0.99 & 24.5 \\
    1.6 & 17 & 124 & 0.80 & 26.4 \\
    \rowcolor{green!25} 1.8 & 14 & 125 & 0.63 & 29.4 \\
    \rowcolor{green!25} 2.0 & 11 & 120 & 0.49 & 34.1 \\ \bottomrule
    \end{tabular}
\end{table}

%% file: evaluation.tex
\section{Evaluation of our Methodology}

\par To demonstrate the cross-domain efficacy of our methodology, we evaluate its performance across four distinct benchmarks: ResNet-18 on CIFAR-10\protect\footnote{Initial Model was adopted from \url{https://github.com/Xilinx/brevitas/tree/master/src/brevitas_examples/imagenet_classification/a2q}}, ResNet-50, and ViT-B/32 for ImageNet classification, and the OPT-125M transformer on the WikiText-2 dataset. In our setup, we define a rigorous optimization constraint where the total accuracy degradation must remain below 1\% Top-1 Accuracy for the vision tasks and within 1 point of Perplexity for the language model relative to the 8-bit baseline, and execute 200 refinement iterations for each model. 

\subsection{Accumulation-Aware Optimization Results}

\par The impact of our iterative optimization loop on the benchmarks' numerical and hardware profiles is illustrated in Figure \ref{fig:optim_results}. Initially, the voltage-scaled baseline (INT8, 0.5V @ 1GHz) exhibits high numerical "pressure," with Pmin distributions statistically centered near 19 bits and a 2.0$\times$ latency penalty, resulting in a suboptimal relative EDP increase of 45\% compared to the (INT8, 0.7V @ 2GHz) solution. By only applying MPQ and unstructured pruning, we effectively truncate the dynamic range of partial products. A key observation here is the strong correlation between the two optimization knobs: lower-precision operands directly enable pushing the accumulator width to lower boundaries, even before the application of AAT. \AM{By utilizing this dynamic range control along our modified architecture, we yield energy gains of up to 32\% while limiting the latency penalty to 47\%, a substantial improvement over the 100\% penalty required by a traditional 0.5V "safe-mode" baseline.}

\par The integration of AAT further compresses $P_{min}$ into power-optimized operating regions, yielding an additional 7\% energy gain beyond the pruning phase. However, the layer-wise breakdown in Figure \ref{fig:optim_results} (bottom) reveals distinct sensitivities with respect to DNN architectures. While CNNs adapt readily to reduced bit-widths, the ViT-B32 and OPT-125M models necessitate higher precision (W6 and W8) across many transformer blocks to maintain accuracy within 1\%. This requirement limits the frequency-scaling headroom, forcing many layers to operate at 1.4--1.8 GHz with higher accumulator bit-widths. Consequently, these models exhibit more moderate EDP reduction (0.80 for ViT and 0.83 for OPT) compared to the CNN architectures due to the sustained latency overhead. This distribution underscores the necessity of a heterogeneous $P_{min}$ strategy. By adaptively assigning precision based on layer sensitivity, the framework ensures rapid convergence and accuracy while reclaiming the performance-energy slack lost to process variability, enabling an average EDP reduction of 23\% across all benchmarks.

\subsection{Comparison against State-of-the-Art}

\par We evaluate our co-optimized design against DNN accelerators that employ aggressive voltage scaling as their primary optimization strategy. To mitigate the circuit and power discrepancies inherent in different technology nodes, we report the relative energy savings of each technique with respect to its own nominal-voltage baseline. Table \ref{tab:sota_comparison} summarizes these energy savings alongside the respective technology and benchmark (DNN model and dataset). X-TPU \cite{xtpu} applies selective voltage scaling to individual multiplier units within a systolic array based on the predicted importance of a neuron’s output, achieving 35\% energy savings with under 1\% accuracy loss. Similarly, \cite{mac_vt} utilizes a systolic array with heterogeneous PEs of varying drive strengths, proposing an optimal mapping of sensitive weights to robust MAC units to achieve 34\% savings. X-NVDLA \cite{xnvdla} extends the NVDLA architecture by applying undervolting specifically to components that propagate to the LSBs of accumulated results, yielding savings of up to 41.5\%. In the bit-serial domain, GAVINA \cite{gavina} applies varying supply levels depending on bit significance; however, its savings are limited to 20\% due to the throughput bottlenecks of high-precision bit-serial configurations. Finally, ThunderVolt \cite{thundervolt_opt} deviates from purely preventative measures by using Razor flip-flops for error detection in a systolic array. It avoids re-execution penalties by dropping erroneous MAC operations and utilizing the spare cycles for correction, yielding energy savings of 45\% on AlexNet.

\input{Tables/sota_comp}

\par Compared to these state-of-the-art techniques, \AM{our post-PnR evaluation in an advanced 2nm technology node demonstrates} a superior balance in energy consumption and throughput, while also accounting for the effects of process variability, associated with low voltage (this issue is explicitly acknowledged only by ThunderVolt \cite{thundervolt_opt}). As shown in Table \ref{tab:sota_comparison}, our methodology yields energy savings of 41\% and 43\% for ResNet-18 and ResNet-50, respectively, when compared to the 0.7V nominal baseline. Furthermore, when compared to a traditional "safe-mode" 0.5V implementation, which would require a 50\% frequency reduction to guarantee deterministic reliability, our approach provides additional energy gains of 18\% to 20\%. Finally, \AM{in contrast to the previous works,} we evaluate the versatility of our framework on contemporary DNN architectures, such as Vision Transformers and small-scale Language Models, demonstrating energy gains of up to 38\% for ViT-B/32 and 37\% for OPT-125M compared to the nominal 0.7V baseline (2GHz), while strictly adhering to the established predictive performance constraints. \AM{Compared to the "safe-mode" 0.5V baseline (1GHz), the energy gains for these models are more modest (14\% for ViT and 12\% for OPT-125M); however, the EDP vastly decreases by over 42\% across these two benchmarks, reflecting the significant throughput recovery enabled by our frequency-scaling strategy.}


%% file: Tables/sota_comp.tex
\begin{table}[t]
\centering
\caption{Comparison with State-of-the-Art Voltage-Scaling DNN Accelerators under 1\% accuracy constraint.}
\label{tab:sota_comparison}
\resizebox{\columnwidth}{!}{%
\begin{tabular}{lcccc}
\toprule

\textbf{Work} & \textbf{Node} & \textbf{VDD Range} & \textbf{Benchmark$^{\dagger}$} & \textbf{Energy Gains} \\ \midrule

X-TPU \cite{xtpu} & 15nm & 0.5V - 0.8V & ResNet-50 (C) & 35.0\% \\

JETCAS '19$^{\ddagger}$  \cite{mac_vt} & 65nm & 0.73V - 1.08V & ResNet-50 (I) & 34.0\% \\

X-NVDLA \cite{xnvdla} & 15nm & 0.4V - 0.8V & ResNet-50 (I) & 41.5\% \\

GAVINA \cite{gavina} & 12nm & 0.35V - 0.55V & ResNet-18 (C) & 20.0\% \\

ThunderVolt$^{\ddagger}$  \cite{thundervolt_opt} & 45nm & 0.65 - 1.0 $^{\S}$ & AlexNet (I) & 45.0\% \\ \midrule

\textbf{This Work} & \textbf{2nm} & \textbf{0.5V} & \textbf{\makecell[l]{ResNet-18 (C) \\ ResNet-50 (I)}} & \textbf{\makecell[c]{41\% / 43\%*}} \\ \bottomrule

\end{tabular}%
}

\begin{flushleft}
\small 
$^*$Reported gains are relative to 0.7V @ 2GHz. When compared to 0.5V @ 1GHz baseline, energy gains are 18\% and 20\% respectively. \\
$^{\dagger}$(I): ImageNet | (C): CIFAR-10. |  $^{\S}$ Refers to Vdd ratio\\
$^{\ddagger}$Contribution of SRAM is not considered in power measurements.
\end{flushleft}
\end{table} 

%% file: conclusions.tex
\section{Conclusions}

\par In this work, we presented a variation-aware methodology targeting systolic architectures in advanced technology nodes. By characterizing the timing-error profiles at low supply voltages, we identified high-magnitude MSB transients as the primary barrier to reliable inference. Our proposed hardware-software solution combines MSB-masking with Accumulation-Aware Quantization to enforce numerical compliance without sacrificing throughput. Evaluations across several DNNs demonstrate that our approach maintains predictive integrity within 1\% of the 8-bit baseline while delivering 39\% energy savings over 0.7V nominal operation and a 46\% EDP improvement over the "always safe" low-Vdd baseline.